\documentclass[letter]{aa} % for the letters 
\usepackage{txfonts}
\usepackage{graphicx}
\usepackage{rotating}
\usepackage{natbib}
\usepackage{lscape}

\bibpunct{(}{)}{;}{a}{}{,} %to follow the A&A style

\newcommand{\ha}{H$\alpha$}
\newcommand{\brg}{Br$\gamma$}
\newcommand{\pab}{Pa$\beta$}
\newcommand{\htwo}{H$_2$}
\newcommand{\hi}{\ion{H}{i}}

\newcommand{\feii}{[\ion{Fe}{ii}]}
\newcommand{\sii}{[\ion{S}{ii}]}
\newcommand{\hei}{\ion{He}{i}}
\newcommand{\caii}{\ion{Ca}{ii}}
\newcommand{\fcaii}{[\ion{Ca}{ii}]}
\newcommand{\oi}{[\ion{O}{i}]}
\newcommand{\noi}{\ion{O}{i}}

\newcommand{\av}{$A_V$}

\newcommand{\um}{$\mu$m}

\newcommand{\msun}{M$_{\odot}$}
\newcommand{\rsun}{R$_{\odot}$}
\newcommand{\msunyr}{M$_{\odot}$\,yr$^{-1}$}
\newcommand{\macc}{$\dot{M}_{acc}$}
\newcommand{\mloss}{$\dot{M}_{loss}$}

\newcommand{\lacc}{$L_{\mathrm{acc}}$}
\newcommand{\rstar}{$R_{\mathrm{*}}$}

\newcommand{\mstar}{$M_{\mathrm{*}}$}

\begin{document}
\title{Recent outburst of the young star V1180 Cas}
%   \subtitle{}
\author{S. Antoniucci$^1$, A.A. Arkharov$^{2}$, A. Di Paola$^1$, T. Giannini$^1$, A. Harutyunyan$^{3}$, E.N. Kopatskaya$^{4}$, V.M. Larionov$^{2,4}$, G. Li Causi$^{1,5}$, D. Lorenzetti$^1$, D. Morozova$^{4}$, B. Nisini$^{1}$, F. Vitali$^1$}

\institute{$^1$ INAF-Osservatorio Astronomico di Roma, via Frascati 33, I-00040 Monte Porzio Catone, Italy\label{oar}\\
$^2$ Central Astronomical Observatory of Pulkovo, Pulkovskoe shosse 65, 196140 St.Petersburg, Russia\label{pulkovo}\\
$^3$ Fundaci\'on Galileo Galilei - INAF, Telescopio Nazionale Galileo, 38700 Santa Cruz de la Palma, Tenerife, Spain\label{tng}\\
$^4$ Astronomical Institute of St.Petersburg University, Russia\label{univ}\\
$^5$ INAF-Istituto di Astrofisica e Planetologia Spaziali, via Fosso del Cavaliere 100, 00133 Roma, Italy\label{iaps}\\
}
\date{}

\abstract
{}
{We report on the ongoing outburst of the young variable V1180 Cas, which is known to display characteristics in common with EXor eruptive variables. We present results that support the scenario of an accretion-driven nature of the brightness variations of the object and provide the first evidence of jet structures around the source.}
{We monitored the recent flux variations of the target in the $R_C$, $J$, $H$, and $K$ bands. New optical and near-IR spectra taken during the current high state of V1180 Cas are presented, in conjunction with \htwo\ narrow-band imaging of the source.}
{Observed near-IR colour variations are analogous to those observed in EXors and consistent with excess emission originating from an accretion event. The spectra show numerous emission lines, which indicates accretion, ejection of matter, and an active disc. Using optical and near-IR emission features we derive a mass accretion rate of $\sim3\times 10^{-8}$\msunyr, which is an order of magnitude lower than previous estimates. In addition, a mass loss rate of $\sim 4\times 10^{-9}$ and $\sim 4\times 10^{-10}$\msunyr\ are estimated from atomic forbidden lines and \htwo, respectively.
Our \htwo\ imaging reveals two bright knots of emission around the source and the nearby optically invisible star V1180 Cas B, clearly indicative of mass-loss phenomena. Higher resolution observations of the detected jet will help to clarify whether V1180 Cas is the driving source and to determine the relation between the observed knots.}
{} 
\authorrunning{S.Antoniucci et al.}
%\titlerunning{}

%Fig.~1 is also available in electronic form
%at the CDS via anonymous ftp to cdsarc.u-strasbg.fr (130.79.128.5)
%or via http://cdsweb.u-strasbg.fr/cgi-bin/qcat?J/A+A/

\maketitle
%
%________________________________________________________________
\section{Introduction}{\label{sec:sec1}
Variability is a common feature among young stellar objects (YSOs). Detected flux variations 
can be generally classified as periodic or recurrent (without any definite period). Hot or cold spots on a rotating stellar surface or a systematic obscuration by an intervening body belong to the first class, whereas episodic mass-accretion variations or the occurrence of unpredicted extinction in front of the YSO are typical of the second. To ascertain whether a non-periodic variation is accretion-driven is of great importance because accretion outbursts can substantially affect the circumstellar disc structure and evolution (disc fragmentation and planet formation). 
To investigate these flux variation modalities, we have recently started an observational programme dubbed EXORCISM \citep[EXOR OptiCal and Infrared Systematic Monitoring,][]{antoniucci13d} to photometrically and spectrally monitor about 20 objects in the 0.5-2.5 $\mu$m range that are identified as known eruptive variables (EXor) or candidates. We recall that EXors 
are pre-main sequence sources that are characterised by outburst events (with typical flux variations of 3-4 mag) that last one year or shorter, with a recurrence time of months to years and present emission line spectra \citep[e.g.][]{herbig08,lorenzetti09,audard14}. One of the targets of the EXORCISM project is V1180 Cas, 
%($\alpha_{2000}$ = 20$^{h}$45$^{m}$53.96$^{s}$, $\delta_{2000}$ =  +67$^{\circ}$57$^{\prime}$38.9$^{\prime \prime}$), 
a young source that belongs to the L1340 star-forming cloud; it was first recognised to be a H$\alpha$ emitter by \citet{kun94} and was identified as [KOS94] HA11. The source is characterised by strong brightness variations ($\Delta I \sim 6$ mag) and was studied by \citet[][K11]{kun11}, who concluded that it is an accretion-driven object although a certain periodicity recognisable in its light curve ($I_C$ band) might indicate extinction-driven dips. 
Few months ago V1180 Cas has entered a new high-brightness phase, which we have been monitoring \citep{antoniucci13c}. We have collected evidence that the luminosity increase is due to an enhanced accretion event, in agreement with the interpretation of K11.
In this Letter, we present new optical/near-IR spectroscopy and photometry of V1180 Cas in the outburst phase, together with imaging in the narrow-band \htwo\ 2.12\um\ filter, providing first evidence of a jet around the eruptive variable.
We also derive estimates for the mass accretion rate (\macc) and correlated mass-loss rate (\mloss), and compare them with
previously obtained values.

\vspace*{-0.1cm}
\section{Observations}{\label{sec:obs}

Optical photometric observations in the $R$ band were obtained at the 70cm AZT-8 telescope of the Crimean Observatory (Ukraine) with a photometer based on ST-7XME SBIG CCDs that is equipped with an $R_C$ filter.
Near-IR photometry with the $JHK$ broad-band filters was taken at the 1.1m AZT-24 telescope located at Campo Imperatore (L'Aquila, Italy), with the imager/spectrometer SWIRCAM \citep{dalessio00}. 
Frames were acquired using a dithering technique and were reduced through standard procedures for bad-pixel removal,
dark correction (optical), flat-fielding, and sky subtraction (near-IR).
Derived optical and near-IR magnitudes are given in Table~\ref{mag:tab}.
% dire del fero?
Additional imaging in the narrow-band filter centred on the 
%\feii(1.64\um) and 
\htwo\ 2.12\um line (and ancillary imaging in $K$)
%and $H$
was acquired with SWIRCAM (pixel scale 1.04\arcsec/pix) on 2014 March 1 and 2 (JD 2456718-9).
Given the similar seeing conditions on the two nights, all good frames were combined to obtain the final \htwo\ and $K$-band images (see Sect.~\ref{sec:jet}),
which correspond to a total integration time of 30 and 5 minutes, respectively.
%shown in Fig.~\ref{h2:fig},

%One $JHK$ photometry set (taken on JD 24056614.5) was instead acquired at TNG (see below).
%Imaging in the narrow-band \htwo\ 2.12\um\ filter was carried out with the AZT-24 Campo Imperatore Telescope on 1, 2, 7, and 21 Mar 2014.
%All images were reduced with the standard procedures for bad pixel removal,
%flat fielding, and sky subtraction, then combined to obtain a final image corresponding to a total integration time of ... 

Low-resolution long-slit spectroscopy covering the interval 0.5--2.4 $\mu$m was obtained
at Telescopio Nazionale Galileo (TNG), La Palma (Canary Islands).
The optical spectrum was taken on 2014 February 19 (JD 2456707) using the DOLORES spectro-imager with the LR-R grism ($\mathcal{R}$ $\sim$ 700) for a total integration time of 15 minutes.
The near-IR segment was acquired with the NICS spectro-imager on 2013 November 18 (JD 2456614) equipped with the $IJ$ and $HK$ grisms ($\mathcal{R}$ $\sim$ 500), by using
the standard ABB$\arcmin$A$\arcmin$ technique and integrating for 40 minutes in $IJ$ and 16 minutes in $HK$. 
All spectral images were flat-fielded, sky-subtracted, and corrected 
for optical distortions. Telluric lines were removed only in the near-IR bands by dividing the extracted spectra by that 
of a normalised telluric standard star after correcting for its intrinsic spectral features. 
Wavelength calibration for the three grisms was achieved 
from reference-lamp exposures. Flux calibration was instead
obtained from additional photometry taken on the same night
as the spectra (see Table~\ref{mag:tab}), in $JHK$ with NICS and at optical wavelengths with DOLORES ($B=20.37$, $V=18.65$, $R=16.94$, $I=15.38$, with uncertainties of 0.03 mag).

%%%%   TABLE - MAG  %%%%%%%%%%%%%%%%%%%%%%%%%%%%%%%%%%%%%%%%%%%%%%%%
\begin{table}
\caption[]{$R_C$ and near-IR photometry and colours of V1180 Cas. Typical errors are lower than 0.03 mag.
    \label{mag:tab}}
\begin{scriptsize}
\begin{tabular}{lcccccc}
\hline
JD-2450000.5     &  $R_C$   & $J$   & $H$   & $K$   & J-H  & H-K   \\
\hline
\hline
6506            &  ...   & 15.64 & 13.69 & 12.06 & 1.95 & 1.63  \\
6510            &  ...   & 14.43 & 12.90 & 11.66 & 1.53 & 1.24  \\
6511            &  ...   & 14.68 & 13.08 & 11.77 & 1.60 & 1.31  \\
6515            &  ...   & 15.65 & 13.65 & 11.96 & 2.00 & 1.69  \\
6517            &  ...   & 15.68 & 13.65 & 12.03 & 2.03 & 1.62  \\
6520            &  ...   & 13.92 & 12.60 & 11.53 & 1.32 & 1.07  \\
6539            &  ...   & 16.26 & 14.00 &  ...  & 2.26 &  ...  \\
6540            &  ...   & 15.95 & 13.74 & 12.00 & 2.21 & 1.74  \\
6558            &  ...   & 13.60 & 12.21 & 11.13 & 1.39 & 1.08  \\
6559            &  ...   & 13.60 & 12.23 & 11.14 & 1.37 & 1.09  \\
6561            &  ...   & 13.50 & 12.15 & 11.04 & 1.35 & 1.11  \\
6562            & 17.00  & 13.50 & 12.17 & 11.04 & 1.33 & 1.13  \\
6563            & 16.96  &  ...  &  ...  &  ...  &  ... &  ...  \\
6565            & 16.98  &  ...  &  ...  &  ...  &  ... &  ...  \\
6575            & 16.74  & 13.48 & 12.06 & 10.82 & 1.42 & 1.24  \\
6576            & 16.96  &  ...  &  ...  &  ...  &  ... &  ...  \\
6577            & ...    & 13.55 & 12.18 &  ...  & 1.37 &  ...  \\
6578            & ...    & 13.56 & 12.13 & 10.95 & 1.43 & 1.18  \\
6579            & ...    & 13.66 & 12.19 & 11.04 & 1.47 & 1.15  \\
6590            & ...    & 13.51 & 12.10 & 11.01 & 1.41 & 1.09  \\
6591            & ...    & 13.51 & 12.12 & 11.03 & 1.39 & 1.09  \\
6595            & ...    & 13.46 & 12.11 & 11.07 & 1.35 & 1.04  \\
6614$^a$        & ...    & 13.46 & 12.21 & 11.13 & 1.25 & 1.08  \\
6644            & ...    & 14.02 & 12.40 & 11.13 & 1.62 & 1.27  \\
6650            & ...    & 13.88 & 12.35 & 11.14 & 1.53 & 1.21  \\
6686            & 17.20  &  ...  &  ...  &  ...  &  ... &  ...  \\
6707$^a$        & 16.94  &  ...  &  ...  &  ...  &  ... &  ...  \\
6718            & ...    & 13.37 & 12.04 & 10.97 & 1.33 & 1.07  \\
6719            & ...    & 13.46 & 12.14 & 11.04 & 1.32 & 1.10  \\
6724            & ...    & 13.44 & 12.13 & 11.09 & 1.31 & 1.04  \\
6737            & ...    & 13.42 & 11.97 & 10.94 & 1.45 & 1.03  \\
6738            & ...    & 13.32 & 11.95 & 10.91 & 1.37 & 1.04  \\
\hline
\end{tabular}
\end{scriptsize}
\\~
$^a$: Photometry taken at TNG.  
\end{table}
%%%%%%%%%%%%%%%%%%%%%%%%%%%%%%%%%%%%%%%%%%%%%%%%%%%%%%%%%%%%%%%%%%%%%%%%
%
%%%%%%%%%%%%%%%%%%%%%%%%%%%%%%%%%%%%%%%%%%%   figure 1 %%%%%%%%%%%%%%%%%%%%%%%%%%%%%
%\begin{figure}[!]
%\centering
%\includegraphics[width=7.5cm]{light_our.eps}
%\caption{V1180 Cas $R_C$ and near-IR ($JHK$) light-curves.\label{lightcurve:fig}}
%\end{figure}
%%%%%%%%%%%%%%%%%%%%%%%%%%%%%%%%%%%%%%%%%%%%%%%%%%%%%%%%%%%%%%%%%%%%%%%%
%%%%%%%%%%%%%%%%%%%%%%%%%%%%%%%%%%%%%%%%%%   figure 2 %%%%%%%%%%%%%%%%%%%%%%%%%%%%%
\begin{figure}[!t]
\centering
\vspace{-0.8cm}
\includegraphics[width=7.0cm]{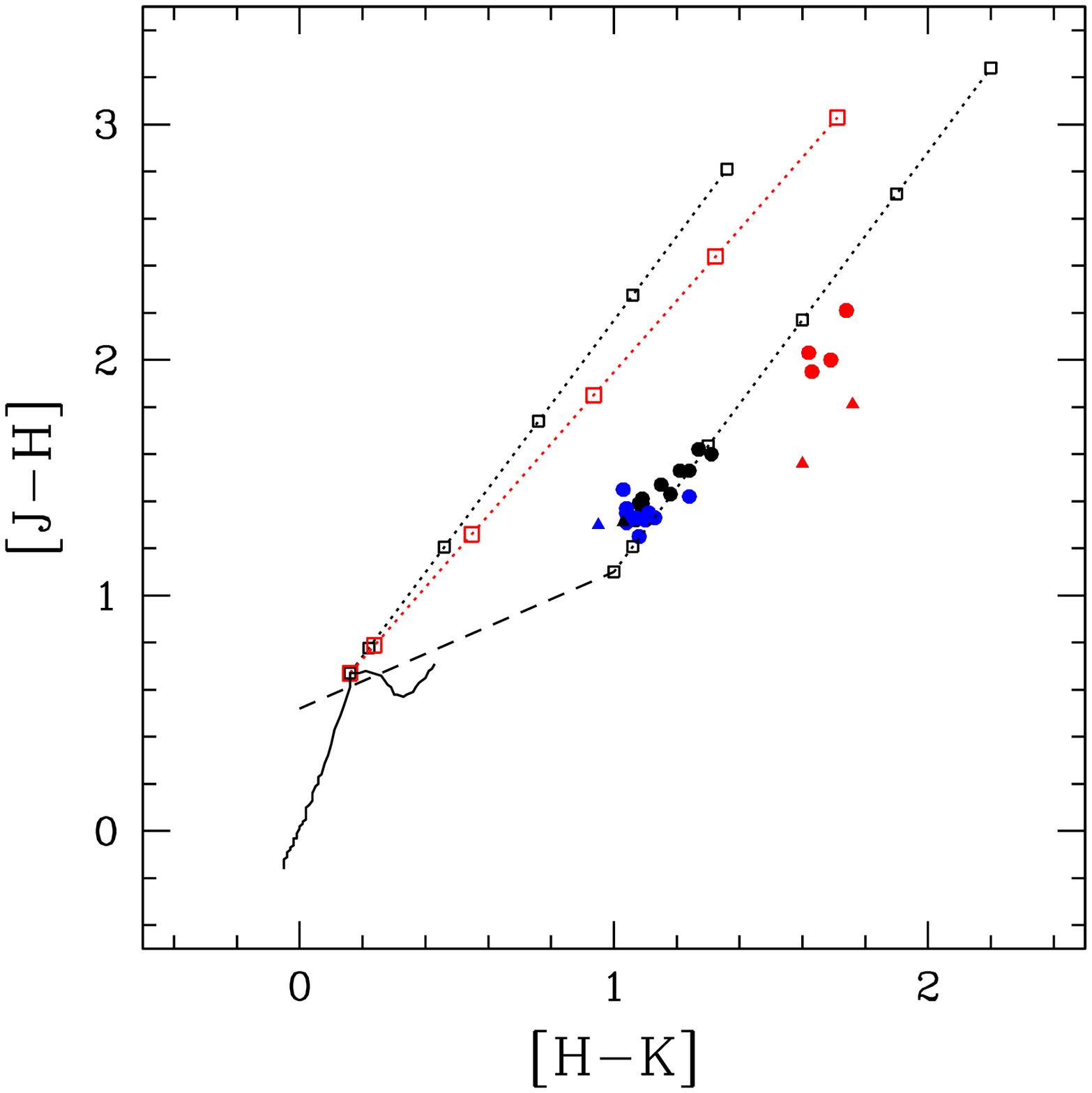}
\caption{V1180 Cas near-IR two-colour plot. Points with $J<$13.50 mag and $J>$15.5 mag are marked in blue and red, respectively. 
Circles refer to our data, triangles show results form 2MASS \citep{2mass} and K11.
The solid line shows colours of main-sequence stars, the dashed line the locus of T Tauri stars \citep{meyer97}. 
The dotted lines indicate the extinction vectors from \citet{rieke85} (black) and \citet{cardelli89} (red), with squares marking the positions corresponding to \av\ = 0, 1, 5, 10, 15, and 20 mag.
%Nel testo puoi scrivere che blue sono quelle con J<13.50 a red quell con J>15.5 
\label{nircol:fig}}
\end{figure}
%%%%%%%%%%%%%%%%%%%%%%%%%%%%%%%%%%%%%%%%%%%%%%%%%%%%%%%%%%%%%%%%%%%%%%%
%%%%%%%%%%%%%%%%%%%%%%%%%%%%%%%%%%%%%%%%%%   figure 3 %%%%%%%%%%%%%%%%%%%%%%%%%%%%%
\begin{figure*}[]
\centering
\includegraphics[width=15.0cm]{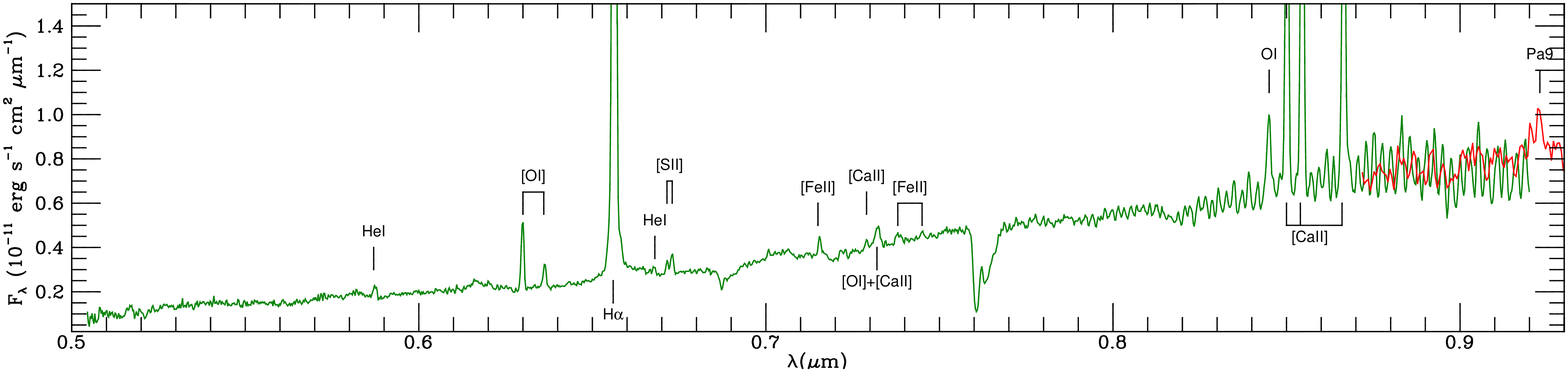}\\
\includegraphics[width=15.0cm]{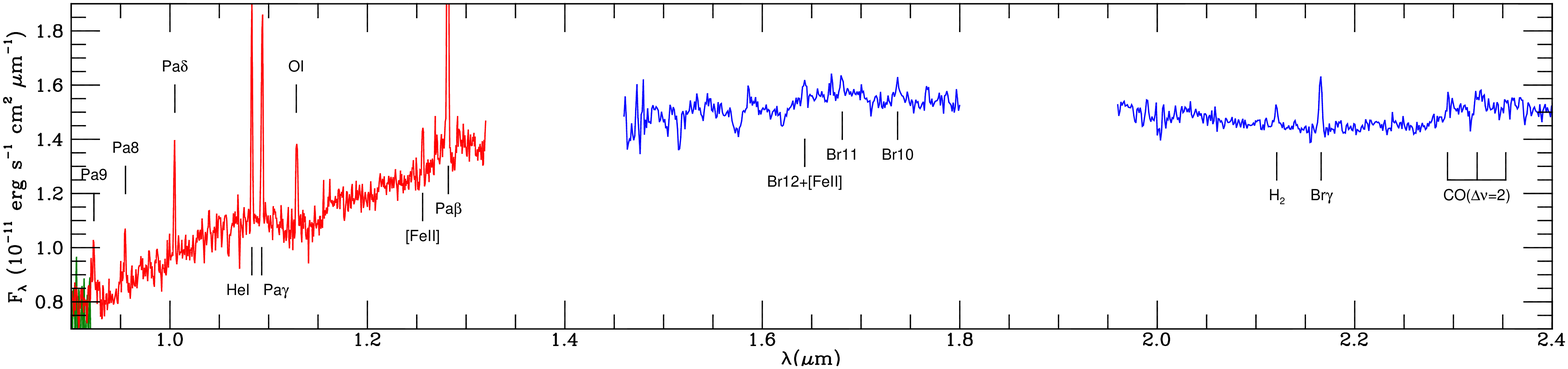}
\caption{\label{spectrum:fig} V1180 Cas optical (top) and near-IR (bottom) spectrum with main emission features labelled.
The spectrum is also available in electronic form at the CDS via anonymous ftp to cdsarc.u-strasbg.fr (130.79.128.5) or via http://cdsweb.u-strasbg.fr/cgi-bin/qcat?J/A+A/.
} 
\end{figure*}
%%%%%%%%%%%%%%%%%%%%%%%%%%%%%%%%%%%%%%%%%%%%%%%%%%%%%%%%%%%%%%%%%%%%%%%
%

\section{Results and discussion}
\subsection{Brightness and colour variations}{\label{sec:var}

Our measurements cover a short period of about 230 days, starting about 2.5 yr after the light-curve presented
by K11 (their Fig.~2).
The main outburst event ($\Delta$J $\simeq$ 2.4, $\Delta$H $\simeq$ 1.7, $\Delta$K $\simeq$ 1.0 mag) was first detected on JD 2456558 (2013-Sep-24).
During the preceding two months, the target also showed some rapid and less pronounced flux variations (up to $\Delta$J $\simeq$ 1.5, $\Delta$H $\simeq$ 1.0, and $\Delta$K $\simeq$ 0.5 mag on JD 2456520).
Since the main outburst event the source has approximately remained at the same brightness level. 
The high state of V1180 Cas has therefore persisted for 180 days, which is 
definitely similar in duration with the long high states monitored by K11. 
Based on our photometry, the brightness level of the current event is approximately the same as in the previous high state between 2008 and 2010. 
We note that the $I_C$-band light-curve of K11 presents a few minima (with typical durations of hundreds of days) that are not periodic, but seem somehow recurrent. During our short monitoring we have not yet detected such long minima, therefore we cannot confirm whether they recur.

Our near-IR data represent the most numerous sample collected so far. They can be used to construct the first meaningful two-colour [$J-H$], [$H-K$] plot for V1180 Cas. This is depicted in Fig.~\ref{nircol:fig} by using both our data (solid circles) and those from 2MASS \citep{2mass} and K11 (solid triangles). 
The IR colours corresponding to the highest ($J \leq$ 13.5 mag) and lowest ($J \geq$ 15.5 mag) states are given in blue and red. As expected for accretion-driven bursts \citep{lorenzetti12a}, the source becomes bluer (redder) while brightening (fading).
Although the near-IR colour changes might be formally compatible with those produced by extinction, especially considering the \citet{cardelli89} law (see Fig.~\ref{nircol:fig}), this hypothesis is ruled out because it would imply an \av\ variation of about 10 mag. This is totally incompatible with the optical colour fluctuations, which are consistent with a $\Delta$\av$\sim$ 1 (see Fig.~2 in K11). %
%, probably related to dust destruction in the outburst.
Using our DOLORES photometric dataset, we find that the current optical colours $(V-R)$ and $(R-I)$ agree well with those of a K0 star extincted by an \av\ in the range 4--5 mag.
As the near-IR colours of the low state are formally compatible with an \av$\sim$10 mag, 
this indicates a strong near-IR excess in quiescence, suggesting the presence of a massive circumstellar disc.
Although an extinction reduction due to the outburst is possible, it is hard to reconcile this scenario with the optical colour changes observed by K11.

%%%%   TABLE - LINES  %%%%%%%%%%%%%%%%%%%%%%%%%%%%%%%%%%%%%%%%%%%%%%%%

\begin{table}
\caption[]{Emission lines, unreddened fluxes, and derived accretion rates.\label{lines:tab}}
\begin{scriptsize}
\begin{tabular}{lccc}
\hline
Line ID & $\lambda$ &  Flux                                & \macc     \\
     &  (\um)     &  (10$^{-15}$ erg s$^{-1}$ cm$^{-2}$) & (10$^{-8}$\msunyr)     \\
\hline
\hline
%?ID              & 5167   & 0.9  $\pm$ 0.2 &       ...          \\
\hei             & 0.588   & 0.7  $\pm$ 0.2 &        ...         \\
%?ID very wide    & 6174   & 3.0  $\pm$ 0.5 &       ...          \\
\oi              & 0.630   & 3.0  $\pm$ 0.1 &        ...         \\
\oi              & 0.636   & 1.2  $\pm$ 0.1 &        ...         \\
\ha              & 0.656   & 102.0$\pm$ 1.0 &        6  \\
\hei             & 0.668  & 0.3  $\pm$ 0.1 &         2  \\
\sii             & 0.671   & 0.5  $\pm$ 0.1 &        ...         \\
\sii             & 0.673   & 0.9  $\pm$ 0.1 &        ...         \\
%?ID              & 7008   & 0.2  $\pm$ 0.1 &       ...          \\
\feii            & 0.715   & 0.8  $\pm$ 0.1 &        ...         \\
\fcaii           & 0.729   & 0.4  $\pm$ 0.1 &        ...         \\
\oi\ + \fcaii       & 0.732  & 1.7  $\pm$ 0.2 &      ...         \\
\feii            & 0.738   & 0.9  $\pm$ 0.1 &        ...         \\
\feii            & 0.745   & 0.4  $\pm$ 0.1 &        ...         \\
\noi             & 0.844  & 5.0  $\pm$ 0.6 &         3  \\
\caii            & 0.849   & 22.0 $\pm$ 0.8 &        4  \\
\caii            & 0.854   & 24.4 $\pm$ 0.8 &        4  \\
\caii            & 0.866   & 20.0 $\pm$ 0.8 &        4  \\
Pa 9             & 0.923   & 6.3  $\pm$ 0.6 &        4  \\
Pa 8             & 0.955   & 4.8  $\pm$ 0.5 &        2  \\
Pa$\delta$       & 1.005  & 7.1  $\pm$ 0.5 &         1  \\
\hei             & 1.082  & 13.1 $\pm$ 0.5 &         2  \\
Pa$\gamma$       & 1.094  & 15.2 $\pm$ 0.5 &         3  \\
\noi             & 1.128  & 9.0  $\pm$ 0.5 &        ...          \\
\feii            & 1.257  & 4.0  $\pm$ 0.5 &        ...          \\
%ID               & 12686  & 1.3  $\pm$ 0.4 &        ...         \\
\pab             & 1.282  & 35.7 $\pm$ 1.0 &         2  \\
%?ID              & 12911  &      $\pm$     &       ...          \\
%?ID              & 13197  &      $\pm$     &       ...          \\
Br12 + \feii       & 1.642  & 3.6 $\pm$ 0.4  &      ...          \\
Br11             & 1.681  & 2.4 $\pm$ 0.4  &        ...          \\
Br10             & 1.737  & 2.8 $\pm$ 0.4  &        ...          \\
\htwo\ 1--0 S(1)  & 2.121  & 1.4  $\pm$ 0.2 &        ...         \\
\brg             & 2.166  & 7.1  $\pm$ 0.3 &         1  \\
CO(2-0)          & 2.294  & 1.3  $\pm$ 0.2 &        ...          \\
CO(3-1)          & 2.324  & 1.4  $\pm$ 0.2 &        ...          \\
\hline
\end{tabular}
\end{scriptsize}
\end{table}

\vspace*{-0.1cm}
\subsection{Optical and near-IR spectroscopy}{\label{sec:spec}
The flux-calibrated 0.5-2.4 \um\ spectrum of the source is displayed in Fig.~\ref{spectrum:fig}. \ha\ is the brightest feature, with an EW around $-400~\AA$, to be compared with a maximum EW of $-900\AA$ measured by K11 in Aug 2008. Other \hi\ emission lines from the Paschen and Brackett series are also clearly visible. Other permitted transitions are the strong \caii\ triplet (0.85-0.87\um) and the \hei\ line at 1.08\um, which are usually associated with stellar winds \citep[e.g.][]{edwards06}.
We also detected many forbidden emission lines, which are typical tracers of shocks driven by jets \citep[e.g.][]{giannini08}. In particular, we observe \oi\ lines at 0.630 and 0.636~\um, the two \sii\ lines at 0.67 \um, \feii\ emission (0.715, 0.738, 0.745, and 1.25~\um), and the \htwo\ transition at 2.12\um. The simultaneous detection of all these lines testifies to mass-loss phenomena from the source, which is presumably correlated to the ongoing accretion event.
We also report the detection of CO($\Delta\nu=2$) bands in emission longward of 2.295~\um.
CO in emission is usually associated with large amounts of warm gas (T$\sim$3000 K) in the inner regions of the circumstellar disc, which is indicative of an active phase of accretion \citep[e.g.][]{najita96b, lorenzetti09, kospal11}.
The main emission features are listed in Table~\ref{lines:tab}.

We computed the mass accretion rate by using the empirical relationships that connect the line and accretion luminosity (\lacc)
%Refined line-\lacc\ calibrations for many emission lines were 
recently derived by \citet{alcala14} based on VLT/X-Shooter observations of a sample of young T Tauri stars in Lupus.
Adopting the same stellar parameters as in K11 (\mstar\ = 0.8\msun, \rstar\ = 2\rsun), a distance of 600 pc, and a visual extinction \av\ = 4.3 mag, we derive an \macc\ in the range 1--6$\times 10^{-8}$\msunyr\ from 13 emission lines both in the optical and in the near-IR 
(see Table \ref{lines:tab}), with a median value of 3 $\times 10^{-8}$\msunyr.
This accretion rate is about one order of magnitude lower than the one derived by K11 (\macc\ $>$ 1.6$\times 10^{-7}$\msunyr) for the high state of V1180 Cas between 2008 and 2009, using the \caii\ $\lambda 8542$ line and the relative relationship by \citet{dahm08}. 
However, this discrepancy must in part be ascribed to systematics of the employed relationship, because adopting the same Dahm's calibration and the \caii\ $\lambda 8542$ line we obtain \macc\ $\sim$ 1$\times 10^{-7}$\msunyr.
We note that an extinction of \av=3.3 mag can be estimated from the ratio Pa$\delta$/Br$\gamma$, 
assuming that the emission is optically thin \citep[e.g.][]{bary08} and adopting Cardelli's law. This yields a slightly lower median \macc\ value of 2$\times 10^{-8}$\msunyr.
An estimate of the mass-loss flux can be derived from jet emission lines following the procedures described in \citet{antoniucci08}. 
From the \sii\ 0.671/0.673 line ratio we infer a density of 5 10$^3$ cm$^{-3}$, while from the ratio 1.25/1.64~\um\ \feii\ lines, which is theoretically expected to be around 1.1 (Giannini et al., in preparation), we estimate that \av$\sim$0 in the region where these lines originate.
Assuming a gas temperature $T$ = 10\,000 K, a velocity of 100 km/s, and a jet section of 1\arcsec\ (i.e. filling the entire slit width), we obtain an \mloss\ of 3$\times 10^{-9}$ and 5$\times 10^{-9}$ \msunyr\ from the \oi\ 0.630~\um\ and the \feii 1.25~\um\ lines, which is about an order of magnitude lower than \macc, as expected from magneto-hydrodynamical jet-launching models \citep[e.g.][]{ferreira06}.
Using the \htwo\ 2.12~\um\ line, we derive a lower \mloss\ of 4$\times 10^{-10}$ \msunyr\ for the molecular part of the jet.

%
%%%%%%   figure 1 %%%%%%%%%%%%%%%%%%%%%%%%%%%%%
\begin{figure*}
\centering
\includegraphics[width=12.5cm]{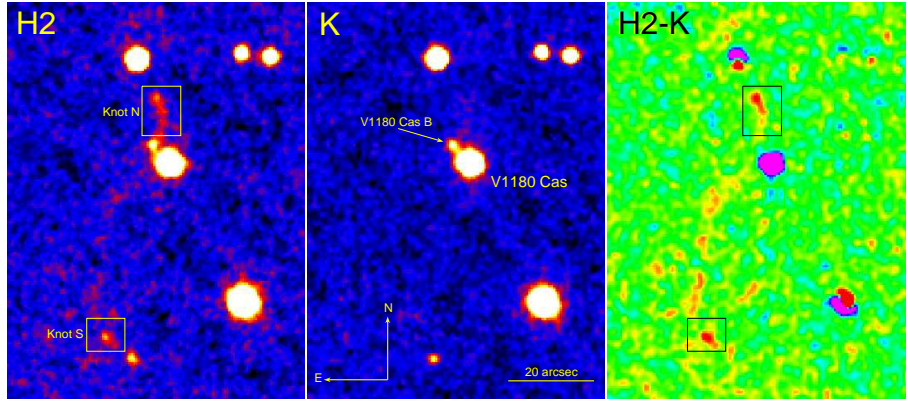}\\
\caption{Images of the region around V1180 Cas taken at Campo Imperatore (a $70\arcsec\times90\arcsec$ field is displayed). The narrow-band \htwo\ image, the $K$-band filter image, and the difference \htwo--$K$ are reported in the left, middle, and right panel. \label{h2:fig}}
\end{figure*}
%%%%%%%%%%%%%%%%%%%%%%%%%%%%%%%%%%%%%%%%
%

\vspace*{-0.1cm}
\subsection{Jet emission}{\label{sec:jet}
The \htwo\ and $K$-band images of the field around V1180 Cas are shown in Fig.~\ref{h2:fig} (left and middle panels). 
The position of V1180 Cas and V1180 Cas B (a star at about 6\arcsec\ NE of the target) are indicated. 
V1180 Cas B is an optically invisible star with a steep spectral energy distribution ($\alpha_{2-24\mu m}=1.9$, K11).
Photometry of V1180 Cas B from our datasets provides steady $K=14.6$ mag and $H>15.1$ mag values, revealing no 
significant variation with respect to 2MASS values ($K=14.7$, $H>16.3$).
The angular distance between the two objects implies a projected separation of 3600 AU, therefore it is not certain that the two stars are physically related.
A direct comparison of the \htwo\ and $K$-band images shows two bright \htwo\ emission regions, one located $\sim$10\arcsec\ north of the target (knot N) and one at $\sim$45\arcsec\ to the south (knot S).
%\textbf{These knots will be named MHO 2964 in the catalogue of Molecular Hydrogen emission-line Objects in outflows from YSOs\footnote{\textbf{http://www.astro.ljmu.ac.uk/MHCat; see \citet{mho}}.} (see Walawender et al., in preparation).}
These knots will are named MHO 2964 in the MHO catalogue\footnote{Catalogue of Molecular Hydrogen emission-line Objects in outflows from young stars \citep{mho}: http://www.astro.ljmu.ac.uk/MHCat.} (see also Walawender et al. in preparation).
The subtraction of the two images (after normalisation to bring the flux count of the stars at the same level) allowed us to remove the continuum emission from the \htwo\ image. Despite some residuals at the position of the bright stars in the field, 
the \htwo--$K$ image (right panel of Fig.~\ref{h2:fig}) shows, in addition to the bright \htwo\ knots N and S, the possible presence of a long (and curved) chain of knots that stretches from the region of the target to knot S.
This morphology is definitely similar to that of other known precessing jets \cite[e.g.][]{lorenzetti02}.
Based on our imaging, it is difficult to understand how knots N and S are related and which is the driving source of the jet. The observed structure suggests that the red star V1180 Cas B might be the driving object, but because of the limited spatial resolution of the images we cannot provide a conclusive answer in this regard.
In any case, inspection of the TNG spectral image reveals no \htwo\ emission outside the object position along the slit, which had been aligned at a P.A.=0\degr.
%In any case, mass loss phenomena occurring at the spatial position of V1180 Cas are definitely testified by the forbidden emission lines detected in our spectra.
Additional observations with higher resolution and sensitivity are definitely required to accurately characterise the detected jet. 
% accennare alle immagini Spitzer?

\vspace*{-0.2cm}
\section{Conclusions}
We presented new optical/near-IR photometry and spectra of the young eruptive variable V1180 Cas,
which is currently in a high-brightness state. Our results support the scenario of an accretion-driven nature of the brightness fluctuations: observed near-IR colour variations are similar to those seen in EXors and the spectra show numerous emission lines that typically trace accretion and the tightly related ejection of matter. Moreover, detection of the CO 2.3~\um\ bands in emission highly suggests large amounts of warm gas in the inner regions of an active accretion disc.

Using 13 optical and near-IR emission features, we derived a mass accretion rate of $3\times 10^{-8}$\msunyr, which is an order of magnitude lower than previous estimates. For the mass-loss rate, we inferred a value of 4 $\times 10^{-9}$ and 4 $\times 10^{-10}$\msunyr\ from atomic forbidden lines (\oi, \feii) and \htwo.

By analysing \htwo\ narrow-band images, we provided the first evidence for a jet in the region of V1180 Cas. The imaging reveals two bright knots of emission around the source and the optically invisible nearby star V1180 Cas B, which are clearly indicative of intense mass loss. 
Higher resolution observations of the jet will help to clarify whether V1180 Cas is the driving source and to determine the relation between the detected knots.

\vspace*{-0.2cm}
%bibtex & natbib----------------------------------------------------------------------
\bibliographystyle{aa} % style aa.bst
\bibliography{refs} % your references Yourfile.bib

% Ciao!
\end{document}